\documentclass[12pt,preprint]{article}
\usepackage{pstricks}
\usepackage{amsmath}
\usepackage{amssymb}
\usepackage{graphicx}
\usepackage{subfig}
\usepackage{float}
\usepackage{cite}
\usepackage{setspace}

\begin{document}

\begin{center}
{\bf \large Large scale anisotropy due to pre-inflationary 
phase of cosmic evolution}

\bigskip
{Pavan K. Aluri and Pankaj Jain}

\bigskip
{Department of Physics, Indian Institute of Technology Kanpur, \\
       Kanpur - 208016, India}
\end{center}

\begin{abstract}
We show that perturbations generated during the anisotropic pre-inflationary
stage of cosmic evolution
may affect cosmological observations today for a certain range of parameters. 
Due to the anisotropic nature of the universe during
such early times, it might explain some of the observed signals of large scale
anisotropy. In particular we argue that the alignment of CMB quadrupole and octopole
may be explained by the Sachs-Wolfe effect due to the large scale anisotropic
modes from very early times of cosmological evolution.
We also comment on how the observed dipole modulation of CMB power may
be explained within this framework.
\end{abstract}

\section{Introduction}
It is generally accepted that at early times, the universe might be anisotropic
and inhomogeneous. The universe may evolve into a de Sitter space-time 
as it expands under the influence of a positive cosmological constant
(Gibbons and Hawking 1977; Hawking and Moss 1982). 
Furthermore it has been shown (Wald 1983) that almost
all Bianchi models asymptotically evolve into a de Sitter space in the presence
of a positive cosmological constant. These are anisotropic but homogeneous models.
The time scale for evolution to isotropy is $\sqrt{3/\Lambda}$, where $\Lambda$
is the cosmological constant during inflation (Wald 1983). 
At time scales $t<\sqrt{3/\Lambda}$, the universe
is anisotropic. Hence the modes which leave the horizon before this time are likely
to show features of anisotropy. 

An interesting observation of breakdown of isotropy in cosmological
data is the alignment of quadrupole and octopole moments of CMBR 
(de Oliveira-Costa {\it et al}
2004; Ralston and Jain 2004; Eriksen {\it et al} 2004; Copi, Huterer and 
Starkman 2004; Schwarz {\it et al} 2004; Land and Magueijo 2005). 
One can define a prefered axis 
for both the quadrupole and octopole (de Oliveira-Costa {\it et al};
Ralston and Jain 2004), which point in the same direction to a 
very good approximation. 
This axis is found to point towards $(l= 237.64^o,b=62.95^o)$, approximately
towards the Virgo cluster of galaxies, 
and makes an angle of approximately $27^o$
with the galactic axis (Aluri {\it et al} 2011).  
A possible explanation of this phenomenon is
foregrounds, systematics or noise (Slosar and Seljak 2004; Abramo, Sodre and
Wuensche 2006; Rakic, Rasanen and Schwarz 2006; Gruppuso, Burigana and
Finelli 2007; Naselsky, Verkhodanav and Nielsen 2008). 
The possibility that foregrounds might cause 
the observed alignment has been ruled out
(Aluri {\it et al} 2011), suggesting that its origin is most likely 
cosmological. It is very interesting that several other cosmological
observations such as radio polarizations (Jain and Ralston 1999),
Optical polarizations (Hutsem\'{e}kers 1998) and the CMB dipole
also points in the same direction (Ralston and Jain 2004). 
Another interesting observation is the dipole modulation of the
CMB power (Eriksen {\it et al} 2004).
 
Some cosmological explanations
for these anomalies include anisotropic inflation
(Berera, Buniy and Kephart 2004; Gordon {\it et al} 2005; Ackerman, Carroll
and Wise 2007; Erickcek, Carroll and Kamionkowski 2008; Kanno {\it et al} 2008;
Yokoyama and Soda 2008; Koivisto and Mota 2008; Boehmer and Mota 2008),
anisotropic/inhomogeneous spaces (Jaffe {\it et al} 2006, Land and Magueijo
2006, Bridges {\it et al} 2007; Ghosh, Hajian and Souradeep 2007; 
Pontzen and Challinor 2007; Kahniashvili, Lavrelashvili and Ratra 2008; 
Carroll, Tseng and Wise 2010) and
local voids (Inoue and Silk 2006).
It has also been suggested  
that initial phase of inflation,
where the kinetic energy of the scalar field dominates, may explain some
of these anomalies (Contaldi {\it et al} 2003; Donoghue, Dutta and Ross, 2009). 
Suitable estimators to characterise these primordial anisotropies
have also been proposed (Hajian, Souradeep and Cornish 2004;
Ralston and Jain 2004; Copi {\it et al} 2006; Bernui {\it et al} 2006; 
Armendiraz-Picon 2006; Pullen and Kamionkowski 2007; Samal {\it et al} 2008; Groeneboom and Eriksen
2009; Bartolo {\it et al} 2011).


A mode with wave number $\vec k$ leaves the horizon at times
\begin{equation}
k|\eta|<1 \,,
\label{eq:cond1}
\end{equation}
where $k=|\vec k|$ and the conformal time $\eta$ is defined as
\begin{equation}
\eta(t) = \int_{t_e}^t {dt'\over a(t')} \,,
\end{equation}
with $t_e$ equal to 
the time at the end of inflation. Note that by this definition
$\eta$ is negative at times before the end of inflation and positive later on.
During inflation, the universe undergoes a rapid phase of expansion with the
scale factor growing as
\begin{equation}
a(t) = a_Ie^{H_It} \,,
\end{equation}
where $H_I$ is the Hubble constant during inflation.
This gives
\begin{equation}
\eta = - {1\over H_I a(t)} \left[1- {a(t)\over a(t_e)}  \right]
        \approx - {1\over H_I a(t)} \,,
\label{eq:eta_before_Inf}
\end{equation}
where we ignore the contribution from $t_e$ since $t_e >> t$.
Throughout this paper we shall assume a spatially flat FRW metric,
after the universe becomes isotropic.

During inflation, the curvature scalar $R=12 H_I^2$. Hence we find that
$\Lambda \sim 12 H_I^2$. This implies that the time, $t_{iso}$, after which
isotropy sets in is of order $t_{iso} \sim 0.5/H_I$. Using Eq. \ref{eq:cond1},
we find that at these early times only modes with wave number,
\begin{equation}
k < H_I a_I e^{0.5} \,,
\end{equation}
leave the horizon. If these modes, which are generated during the
anisotropic phase, before the universe evolves into a de-Sitter space-time, re-enter the horizon before the current era, then 
these could lead to large scale anisotropy in cosmological
observations. 

We may assume that the anisotropy causes all these wave
vectors to lie in a plane and that this plane is perpendicular to the axis
of alignment of the CMB
quadrupole and octopole (de Oliveira-Costa 2004, Ralston and Jain 2004, 
Copi, Huterer and Starkman 2004, Schwarz {\it et al} 2004,
Land and Magueijo 2005). 
We shall refer to this axis as the
prefered axis and the plane perpendicular to it as the prefered plane.
Hence these anisotropic modes will lead to fluctuations in the metric which
lie in a plane. Such anisotropic metric fluctuation could lead, directly or
indirectly, to some of the claimed signals of large scale anisotropy.

The anisotropic metric fluctuations could lead to alignment of the low $l$
CMB multipoles through the Sachs-Wolfe effect 
(Sachs and Wolfe 1967; Hu and Sugiyama 1995;
Francis and Peacock 2010, Dodelson 2003, Durrer 2008, Weinberg 2008). The photons which propagate perpendicular to the prefered plane
do not experience any gravitational potential. Hence these do not get any
contribution due to the Sachs-Wolfe effect. However the photons that propagate
along the plane experience maximum effect since the metric fluctuations are maximal
in this plane. This will lead to additional anisotropy in the CMB temperature
fluctuations. If this effect is sufficiently strong it might lead to
alignment of low $l$ multipoles. We investigate this possibility in the 
present paper.

\section{Anisotropic metric perturbations at large distance scales}
The anisotropic perturbations, which are generated during the early 
anisotropic phase
of cosmic evolution, may re-enter the horizon at late times. A perturbation
that leaves the horizon at conformal time $\eta=-1/k$, re-enters the
horizon at time $\eta=1/k$. In this section we determine the conformal
time at which the primordial perturbations zoomed to super horizon scales
during the early time of inflation, re-enter the horizon.

The conformal time before the end of inflation is given by Eq. \ref{eq:eta_before_Inf}.
At the end of inflation $\eta=0$. We assume that, immediately
after that radiation dominates the energy density of the universe.
The scale factor during this phase may be expressed as,
\begin{equation}
a(t) = A t^{1/2}\,.
\end{equation}
We can fix $A$ by equating this to the inflationary solution at time $t_e$
corresponding to the end of inflation. We find that during the radiation
domination phase,
\begin{equation}
\eta = {2a(t) t_e \over a^2(t_e)} \left[1-{a(t_e) \over a(t)}\right]\,.
\end{equation}
We point out that the time $t_e$ may be fixed by requiring that
the number of e-folds the universe has expanded into, during inflation,
are $N \approx 64$. Hence we require,
\begin{equation}
H_I t_e = N\approx 64\,.
\end{equation}
Hence we find that the conformal time at the radiation matter equality
is equal to,
\begin{equation}
\eta_{eq} \approx {2a_{eq} \over a^2(t_e)} {N\over H_I}\,,
\end{equation}
where $a_{eq} = a(t_{eq})$.

We next compute the conformal time during matter domination phase. We assume
that matter dominates for time $t > t_{eq}$ and ignore all other contributions.
Here we ignore the contribution due to late time acceleration also, since
that will only lead to a small correction to our results. The scale factor
during this phase may be expressed as,
\begin{equation}
a(t) = B t^{2/3}\,.
\end{equation}
We find the conformal time during this phase to be,
\begin{equation}
\eta = 3\, \sqrt{{a(t) a_{eq}\over a^2(t_e)}}\, {N\over H_Ia(t_e)} 
\,\left[1-\sqrt{a_{eq}\over a(t)}\,\right] + \eta_{eq}\,.
\end{equation}
Hence at late times, we find,
\begin{equation}
\eta(t)  \approx  {3N\over H_Ia(t_e)}\, \sqrt{a(t) a_{eq} \over a^2(t_e)} \,.
\end{equation}
Let $t_l$ represent the time when a mode leaves the horizon during
inflation. 
 The time $t_r$, when it re-enters the horizon, is given by,
\begin{equation}
{a(t_e)\over a(t_l)}  \approx  3N{a_{eq}\over a(t_e)}\, \sqrt{a(t_r)\over a_{eq} } \,.
\end{equation}
Here we have assumed that the mode re-enters the horizon during matter
domination.

We next make an estimate of $t_r$ in order to determine whether the
anisotropic modes might have some effect on the cosmological observations
made today. We are interested in computing the time corresponding to
 the modes which leave the horizon
when the universe has just entered the de-Sitter phase. All
the modes which leave the horizon before this time are generated
during the anisotropic phase of evolution. Hence, in our case, the time $t_l$ 
corresponds to the very early time during inflation.
Since inflation lasts approximated 28 efolds, we expect,
\begin{equation}
{a(t_e)\over a(t_l)}  \approx 10^{28}\,.
\end{equation}
Thus we get,
\begin{equation}
{a(t_r)\over a_{eq}} = 7\times 10^{-3} \left[{10^{19} {\rm GeV}\over H_I}\right]
\left[{5\times 10^4 {\rm yr}\over t_{eq}}\right]
\left[{a(t_e)/a(t_l)\over 10^{28}}\right]^2
\left[{64\over N}\right] \,.
\label{eq:atr}
\end{equation}
This equation gives the scale factor, $a(t_r)$, when the modes which left
the horizon when the universe crossed over from anisotropic to de Sitter
phase, re-entered the horizon.
It shows that there exists
allowed parameter range where the anisotropic modes, generated before
the phase of isotropic and homogeneous 
inflation, can re-enter the horizon before the current 
time and hence have observational consequences.
To the best of our knowledge, this simple and basic result, 
Eq. \ref{eq:atr}, 
does not exist in the literature so far. 
We find that if $H_I$ is of the order of the Planck scale then the modes
generated during the anisotropic phase enter the horizon during radiation
dominated phase, i.e. before decoupling. Hence the large wavelength modes
will show signals of statistical anisotropy. These modes may
therefore explain the anomalies observed at low $l$. If $H_I$ is of order
$10^{15}$ GeV these modes enter the horizon during the matter dominated phase.
In this case the anisotropic modes will have no contribution at the time of
decoupling of radiation from matter. However as these anisotropic
modes enter the horizon at late times, they affect the 
low $l$ multipoles through
integrated Sachs-Wolfe (ISW) effect 
(Sachs and Wolfe 1967; Hu and Sugiyama 1995;
Francis and Peacock 2010, Dodelson 2003, Durrer 2008, Weinberg 2008). 

\begin{figure}
\centering
\begin{pspicture}(12,5)
\psline[linecolor=black,linewidth=5pt](0.5,2)(11.5,2)

\psline[linecolor=black,linewidth=3pt](0.5,1.91)(0.5,3)
\psline[linecolor=black,linewidth=3pt](2.0,2)(2.0,3)
\psline[linecolor=black,linewidth=3pt](4.5,2)(4.5,3)
\psline[linecolor=black,linewidth=3pt](8,2)(8,3)
\psline[linecolor=black,linewidth=3pt](11.5,1.91)(11.5,3)

\psline[linecolor=black,linewidth=3pt,linestyle=dashed](3,1)(3,5)
\psline[linecolor=black,linewidth=3pt,linestyle=dashed](7.5,1)(7.5,3)
\psline[linecolor=black,linewidth=3pt,linestyle=dashed](11,1)(11,3)
\psline[linecolor=black,linewidth=3pt,linestyle=dashed]{<->}(7.5,1)(11,1)

\rput(0.5,3.5){{\large $t=0$}}
\rput(2,3.5){{\large $t_i$}}
\rput(4.5,3.5){{\large $t_e$}}
\rput(8,3.5){{\large $t_{dec}$}}
\rput(11.5,3.5){{\large $t_0$}}

\rput(3,0.5){{\large $t_{iso}/t_l$}}
\rput(9.2,0.5){{\large $t_r$}}

\rput(1.7,4.9){{\small Anisotropic}}
\rput(1.7,4.65){{\small universe}}
\rput(5.5,4.9){{\small Homogeneous and isotopic}}
\rput(5.5,4.65){{\small universe}}
\end{pspicture}
\caption{A schematic illustration (not to scale) 
of the various time scales mentioned 
in this paper. The universe evolves anisotropically under the influence
of cosmological constant at time $t_i$. The isotropic inflationary phase
starts beyond the time scale $t_{iso}$. Inflation ends at time $t_e$ and
$t_{dec}$ denotes the time scale of decoupling. For a certain range
of allowed parameters, the anisotropic modes which leave the horizon
before $t_l$ may re-enter the horizon after $t_r$ such that $t_r<t_0$,
where $t_0$ is the current time. Hence these can affect cosmological 
observables.
 }
\label{fig:timeline}
\end{figure}
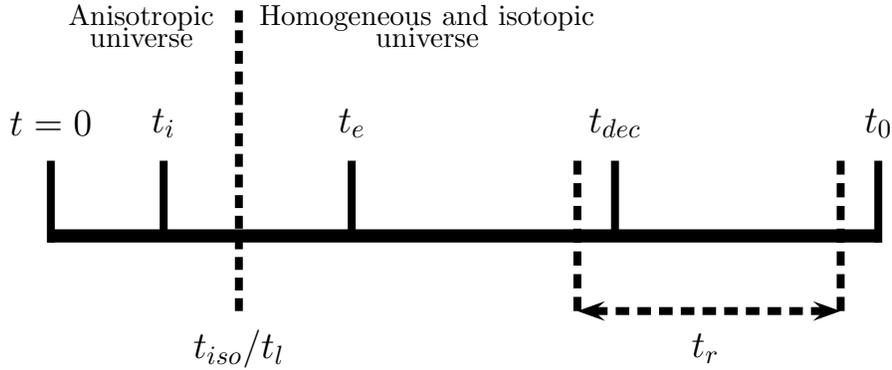

In Fig. [\ref{fig:timeline}] we schematically show a modified cosmic
history, assuming that the parameters of inflation are such the anisotropic
modes enter the horizon before the current era. The corresponding parameter
space is given by Eq. \ref{eq:atr}. In Fig. [\ref{fig:timeline}] the time
$t_i$ corresponds to an early time when the universe starts evolving
under the influence of the cosmological constant. The universe is anisotropic
at this time. It becomes isotropic at time equal to $t_{iso}$. The standard
inflationary phase starts at this time. The anisotropic modes which leave
the horizon at time $t_l$ re-enter the horizon at $t_r$. The precise value
of $t_r$ depends on the inflationary parameters, as governed by Eq. 
\ref{eq:atr}. Hence the anisotropic modes generated before $t_l$ re-enter
the horizon after $t_r$ and can affect cosmological observations if 
$t_r<t_0$, where $t_0$ is the current time.   

We emphasize that if the parameters are such that these anisotropic modes 
re-enter the horizon at a sufficiently early time, then they can in principle
explain the wide range of anisotropic signals observed. These include
the large scale anisotropies seen in CMB (de Oliveira-Costa {\it et al}
2004; Ralston and Jain 2004; Eriksen {\it et al} 2004;
 Copi, Huterer and Starkman 2004; Schwarz {\it et al} 2004), 
optical polarizations (Hutsem\'{e}kers 1998; Agarwal {\it et al} 2011),
radio polarizations (Jain and Ralston 1999), coherent flow in cluster 
peculiar velocities (Kashlinsky {\it et al} 2010) and dipole anisotropy in galaxy
distribution (Itoh {\it et al} 2010). This is due to the wide range of possible anisotropic
models which might be applicable at early times.  
Furthermore the anisotropic nature of the modes created at these 
early times is preserved over much of the history of cosmic evolution.
This is because as long as these modes are outside the horizon, no causal
physics can affect them. Only after they re-enter the horizon do they
start evolving significantly. However at this stage they also start affecting
cosmological observables. 
It would clearly be of great interest to explore a wide range of
these anisotropic models to determine if they can explain the observed 
violations of isotropy.

\section{Low $l$ CMBR multipoles}
As discussed in section 2, we find, using Eq. \ref{eq:atr}, that there
exists a wide range of parameters for which the modes generated during 
the early anisotropic phase of cosmic evolution may play a role in the present
day cosmology. This applies to a wide class of models, discussed in 
Wald (1983), which evolve into de Sitter space-time. 
Due to the large range of allowed models, which fall in this class, it
is possible that a model may exist which might explain the anisotropic
signals observed in CMBR.  
Here we discuss a simple illustrative example of how these anisotropic
modes might lead to the observed low $l$ anomalies. Here we are not interested
in detailed metric models. We shall assume a simple anisotropic model of
perturbations with the prefered direction aligned along the $z$ axis.
Hence we shall assume
 that the modes with wave vectors aligned along the $z$ axis behave
differently from those which lie in the $x-y$ plane. Assuming some
reasonable properties of these modes we determine if they can explain 
alignment of the low $l$ multipoles.

The anisotropic modes leave the horizon at very early times. We assume
that the parameters are chosen such that they re-enter the horizon 
before the time of decoupling. 
As these anisotropic modes re-enter the horizon, they can affect the CMBR
photons. We assume that the wave vectors corresponding to these perturbations
all lie in a plane which is perpendicular to the Virgo axis which is
roughly the axis of alignment of CMBR quadrupole and octopole.
We refer to this axis and the corresponding plane as the preferred axis
and preferred plane respectively. We shall choose our coordinates such
that this anisotropy axis points towards $\hat z$.
Photons propagating perpendicular to this plane would be unaffected by
the anisotropic metric perturbations.
However the photons propagating in other directions would undergo
redshift or blueshift due to the Sachs-Wolfe effect. This effect would
be maximum for photons propagating parallel to the preferred plane.
Hence it would induce additional anisotropies in the CMBR spectrum which
would lie dominantly in the preferred plane. If these have sufficient
strength, then they may yield a preferred axis for quadrupole and octopole,
perpendicular to this plane. Hence this phenomenon can explain the
alignment of $l=2,3$ multipoles.

Let's first assume that these modes, generated during the early 
anisotropic phase
of cosmic evolution, enter the horizon just before decoupling. Let $\psi(\eta_{dec}, \vec x_{dec})$
denote the gravitational potential. Here $\eta_{dec}$ is the conformal time during
decoupling, $\vec x_{dec} = \vec x(\eta_{dec})$ and
\begin{equation}
\vec x(\eta) = \vec x_0 - (\eta_0-\eta) \hat{n}\,,
\end{equation}
where $\vec x_0$ is the position of the observer, $\eta_0$ the conformal time today,
and $\hat n$ is the direction of observation. Ignoring the integrated Sachs-Wolfe
effect we find,
\begin{equation}
\left({\Delta T(\hat n)\over T}\right) \sim \psi(\vec x_{dec},\eta_{dec})\,.
\label{eq:anisotropy}
\end{equation}
We note that here we have assumed the formula corresponding to adiabatic
perturbations. This captures the basic physics that the anisotropies
are related to the metric perturbations. However the detailed final result 
is likely to be dependent on the precise anisotropic model that may
be applicable during the early phase of cosmic evolution. Hence we do not
try to predict the absolute magnitude of these temperature anisotropies.
We assume that by suitable choice of parameters these can be adjusted to
fit the data. Furthermore here we have ignored the contribution due to the
statistically isotropic modes. We assume that these are negligible for
low $l$ and their contribution increases as we increase $l$. Hence for low
$l$ the anisotropic modes dominate whereas these are negligible for 
higher $l$. A more detailed treatment is postponed to future research.

We express the potential in terms of its fourier transform $\tilde\psi(\vec k, \eta)$,
\begin{equation}
\psi(\vec x,\eta) = \int{d^3k \over (2\pi)^3} e^{i\vec k\cdot \vec x} \tilde \psi(\vec k,\eta)\,.
\end{equation}
We next model $\tilde \psi(\vec k,\eta)$ as
\begin{equation}
\tilde \psi(\vec k,\eta) = 2\pi\delta(k_z) g(\vec k_\perp,\eta) \,,
\end{equation}
where $\vec k_\perp$ is the projection of the wave vector in the $x-y$ plane.
This implements our idea that all the modes lie in the $x-y$ plane. It may
of course be useful to replace the delta function by a function strongly
peaked at $k_z=0$. However in the present paper we use only this extreme
model where all the modes lie strictly in the $x-y$ plane.
We denote the magnitude $|\vec k_\perp| = k_\perp$. We assume that
\begin{equation}
g(\vec k_\perp,\eta) \propto 1/k_\perp \,,
\end{equation}
up to a cutoff $k_\perp = k_c$. Beyond this cutoff the anisotropic modes
do not contribute. The anisotropy spectrum generated by these low $\vec k$ modes
may be expressed as,
\begin{equation}
{\Delta T(\hat n)\over T} = {k_c^2\over 2\pi} \int_0^1 y dy J_0(a y) g(k_c y) \,,
\end{equation}
where we have defined the variable $y=k_\perp/k_c$. The symbol $a$ is given by
\begin{equation}
a = \sqrt{{(k_c x_{0\perp})}^2 + {(k_c \Delta\eta)}^2 n_\perp^2 - 2 (k_c \Delta \eta)(k_c\vec x_{0\perp})\cdot \vec n_\perp} \,,
\end{equation}
where $\Delta \eta=\eta_0-\eta_{dec}$ and $\vec x_{0\perp}$ is the component
of the vector $\vec x_0$ in the $x-y$ plane, perpendicular to the prefered direction.

We assume, without loss of generality, that $\vec x_{0\perp}$ points in the
$x$ direction. This simply corresponds to a choice of axes.
If we set this vector to $\vec 0$ then the resulting temperature fluctuations
depend only on $|\vec n_\perp|$, which is uniform in the $x-y$ plane. Hence
in this case the temperature is same every where in the prefered plane. In order
to generate fluctuations in this plane we need another vector such as $\vec x_{0\perp}$.
Alternatively we may assume that the function $g(\vec k_\perp,\eta)$ doesn't
just depend on the magnitude of $\vec k_\perp$. In this case also, we shall
essentially need to introduce another vector in the $x-y$ plane.
In the present paper, we assume that $\vec x_{0\perp}$ is not equal to zero and that
$g(\vec k_\perp,\eta)$ depends only on the magnitude of $\vec k_\perp$.

\begin{figure}
\begin{center}
\includegraphics[width=0.96\textwidth]{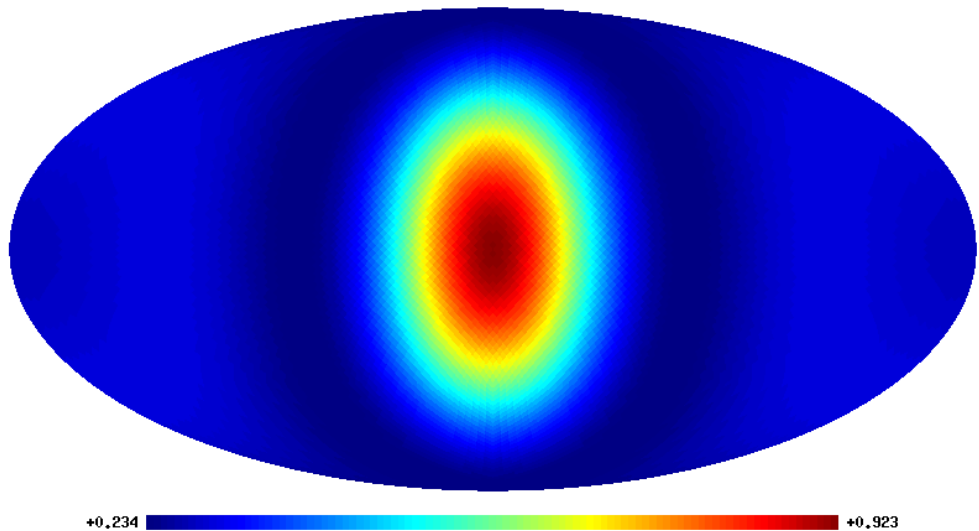}
\end{center}
\caption{The temperature anisotropy generated by
the anisotropic modes due to Sachs-Wolfe effect. The map is generated at
\texttt{HEALPix} resolution of $N_{side} = 32$.}
\label{fig:aniso1}
\end{figure}

In Fig. [\ref{fig:aniso1}], we show the resulting temperature fluctuations, in
arbitrary units, as a function of the position on the sky. Here we have set
$k_c|\vec x_{0\perp}|=6$ and $ k_c\Delta\eta=3$. We speculate that the second
prefered axis in the $x-y$ plane may be the axis of the ecliptic dipolar power
asymmetry (Eriksen {\it et al} 2004). It lies in a plane perpendicular to the 
axis pointing towards Virgo, in approximate agreement with observations. 
Thus, this simple model studied here may simultaneously
account for both the axis of anisotropy seen in the CMB data.

It is clear from the figure that the hot spot of all the modes is aligned
along one direction, which in our coordinate system is the x-axis.
Furthermore if we compute the principal axis (Ralston and Jain 2004, 
Samal {\it et al} 2008), we find that
the axes for $l=2,3,4,5,6$ point towards the z-axis for this map.
The principal eigenvectors (PEV) and the corresponding power is shown
in Table 1.
We find that the power  
due to these anisotropic modes decreases with increase in $l$.  
In fact, for our choice of parameters, the power in multipoles $l\ge 4$ is
negligible compared to the multipoles $l=2,3$.
Hence it is reasonable to assume that  
the multipoles $l\ge 4$ may receive dominant contributions from statistically
isotropic modes. Hence the SW effect considered here may lead to
alignment only for the multipoles $l=2,3$ for our choice of parameters.
It is clear, however, that this is model dependent and results will change
depending on our choice of parameters. Observations show  
 dominant alignment
only between the quadrupole and octopole. However, there is also 
some evidence that
this alignment might continue to larger $l$ (Samal {\it et al} 2008;
Samal {\it et al} 2009).

\begin{table}
\centering

\begin{tabular}{| c | c | c |}
\hline
Multipole, & Power, & PEV \\
$l$ & $l(l+1)C_l/2\pi$ & (x,y,z) \\
  & (in arbitrary units) & \\
\hline
2  &  0.024      &  (  -4.765E-8, \, 1.423E-8, 0.9999) \\
3  &  0.014      &  (  -5.424E-8,   -2.722E-9, 0.9999) \\
4  &  0.0018     &  (  -1.767E-7,   -7.465E-9, 0.9999) \\
5  &  0.00016    &  (  -3.399E-7, \, 4.459E-8, 0.9999) \\
6  &  0.0000058  &  (\, 1.322E-6,   -1.992E-8, 0.9999) \\
\hline
\end{tabular}

\caption{Principal eigenvectors (PEVs) of the CMB signal shown in Fig. 
 [\ref{fig:aniso1}] due to SW effect from 
pre-inflationary anisotropic modes.}
\end{table}

\section{Conclusions}
In this paper, we study the implications of anisotopic primordial
perturbations, carrying imprints of pre-inflationary anisotropic era, 
to the large scale
anomalies found in CMB. We have shown that the modes which leave 
the horizon during the early anisotropic phase 
re-enter the horizon before the current time for a wide range of choice
of the Hubble parameter during inflation. For a certain range of 
allowed values of the Hubble parameter during inflation, these may 
enter the horizon even before decoupling. Hence these  
can provide an explanation of some of the large scale anomalies
of CMB through Sachs-Wolfe effect (or integrated Sachs-Wolfe effect). 
We have described a simple illustrative model of these anisotropic modes
which shows alignment of the low $l$ multipoles. 
For an appropriate
choice of parameters of the model, we have shown that the contributions 
due to the
SW effect is such that they may cause alignment of quadrupole and octopole
in the observed signal. The model requires another prefered axis in the
plane perpendicular to the principal axis of quadrupole and octopole. We
speculate that this second axis might be related to the dipole modulation
axis discovered in (Eriksen {\it et al} 2004).

\bigskip

\textbf{Acknowledgements : }We acknowledge the use of the publicly available
CMBR analysis software \texttt{HEALPix} (Gorski {\it et al} 2005) 
to generate the
images used in the present work. We thank John P. Ralston and Rajib Saha
for useful discussions.

\bigskip
 
{\bf\large  References}

\begin{itemize}

\item[] L. R. Abramo, L. Sodre Jr. and C. A. Wuensche, 2006, Phys. Rev. D, 74, 083515

\item[] L. Ackerman, S. M. Carroll and M. B. Wise, 2007, Phys. Rev. D, 75, 083502

\item[] P. K. Aluri, P. K . Samal, P. Jain and J. P. Ralston, 2011, MNRAS, 414, 1032

\item[] N. Agarwal, P. K. Aluri, P. Jain, P. Tiwari and U. Khanna, 2011,
arXiv:1108.3400

\item[]  C. Armendariz-Picon, 2006, JCAP, 3, 2

\item[] N. Bartolo, E. Dimastrogiovanni, M. Liguori, 
S. Matarrese and A. Riotto, 2011, arXiv:1107.4304 

\item[] A. Berera, R. V. Buniy, and T. W. Kephart, 2004, JCAP, 10, 16

\item[] A. Bernui, T. Villela, C. A. Wuensche, R. Leonardi, I. Ferreira, 2006, A\&A, 454, 409

\item[] C. G. Boehmer and D. F. Mota, 2008, 
       Phys. Lett.  B663, 168

\item[] M. Bridges, J. D. McEwen, A. N. Lasenby and M. P. Hobson, 2007, MNRAS, 377, 1473

\item[] S. M. Carroll, C-Y. Tseng and M. B. Wise, 2010, Phys. Rev. D, 81, 083501

\item[] C. R. Contaldi, M. Peloso, L. Kofman and A. Linde, 2003,
JCAP 0307, 002

\item[] C. J. Copi, D. Huterer and G. D. Starkman, 2004, Phys. Rev. D, 70, 4, 043515

\item[] C. J. Copi, D. Huterer, D. J. Schwarz, G. D. Starkman, 2006, MNRAS, 367, 79

\item[] A. de Oliveira-Costa, M. Tegmark, M. Zaldarriaga and A. Hamilton, 2004, Phys. Rev. D, 69, 063516

\item[] S. Dodelson, \emph{Modern cosmology}, 2003, Academic Press

\item[] J. F. Donoghue, K. Dutta and A. Ross, 
2009, Phys. Rev.  D80 023526

\item[] R. Durrer, \emph{The Cosmic Microwave Background}, 2008, Cambridge University Press

\item[] H. K. Eriksen et. al., 2004, ApJ, 605, 14

\item[] A. L. Erickcek, S. M. Carroll and M. Kamionkowski, 2008, Phys. Rev. D, 78, 083012

\item[] C. L. Francis and J. A. Peacock, 2010, MNRAS, 406, 14

\item[] A. Hajian, T. Souradeep, N. Cornish, 2004, ApJ, 618, L63

\item[] T. Ghosh, A. Hajian and T. Souradeep, 2007, Phys. Rev. D, 75, 083007

\item[] G. W. Gibbons and S. W. Hawking, 1977, Phys. Rev. D 15, 2738

\item[] C. Gordon, W. Hu, D. Huterer and T. Crawford, 2005, Phys. Rev. D, 72, 103002

\item[] K. M. Gorski et. al., 2005, ApJ, 622, 759

\item[] N. E. Groeneboom and H. K. Eriksen, 2009, ApJ, 690, 1807

\item[] A. Gruppuso, C. Burigana and F. Finelli , 2007, MNRAS, 376, 907

\item[] A. E. Gumrukcuoglu, C. R. Contaldi and M. Peloso, 2007, JCAP, 11, 005

\item[] S. W. Hawking and I. G. Moss, Phys. Lett. 110B, 35

\item[] W. Hu and N. Sugiyama, 1995, ApJ, 444, 448

\item[] D. Hutsem\'{e}kers, 1998, Astron. Astrophys., 332, 410

\item[] K. T. Inoue and J. Silk, 2006, ApJ, 648, 23
	
\item[] Y. Itoh, K. Yahata, and Takada, Masahiro, 2010, Phys. Rev. 
D. 82, 043530 

\item[] T. R. Jaffe, S. Hervik, A. J. Banday and K. M. Gorski, 2006, ApJ, 644, 701

\item[] P. Jain and J. P. Ralston, 1999, Mod. Phys. Lett. A 14, 417

\item[] A. Kashlinsky, F. Atrio-Barandela, H. Ebeling, 
A. Edge and D. Kocevski,
2010, Astrophys.J. 712, L81-L85

\item[] T. S. Koivisto and D. F. Mota, 2008, 
   JCAP 0808, 021

\item[] T. Kahniashvili, G. Lavrelashvili and B. Ratra, 2008, Phys. Rev. D, 78, 063012

\item[] S. Kanno, M. Kimura, J. Soda, and S. Yokoyama, 2008, JCAP, 8, 34

\item[] K. Land and J. Magueijo, 2005, Phys. Rev. Lett., 95, 071301

\item[] K. Land and  J. Magueijo, 2006, MNRAS, 367, 1714

\item[] P. D. Naselsky, O. V. Verkhodanov and M. T. B. Nielsen, 2008, Astrophys. Bull., 63, 216

\item[] A. Pontzen and A. Challinor, 2007, MNRAS, 380, 1387

\item[] A. R. Pullen and M. Kamionkowski, 2007, Phys. Rev. D, 76, 103529

\item[] A. Rakic, S. Rasanen and D. J. Schwarz, 2006, MNRAS, 369, L27

\item[] J. P. Ralston and P. Jain, 2004, Int. J. Mod. Phys. D, 13, 1857

\item[] R. K. Sachs and A. M. Wolfe, 1967, ApJS, 147, 73

\item[] P. K. Samal, R. Saha, P. Jain and J. P. Ralston, 2008, MNRAS, 385, 1718

\item[] P. K. Samal, R. Saha, P. Jain and J. P. Ralston, 2009, 
MNRAS, 396, 511 

\item[] D. J. Schwarz, G. D. Starkman, D. Huterer and C. J. Copi, 2004,  Phys. Rev. Lett., 93, 221301

\item[] A. Slosar and U. Seljak, 2004, Phys. Rev. D, 70, 083002

\item[] R. M. Wald, 1983, Phys. Rev. D, 28, 2118R

\item[] S. Weinberg, \emph{Cosmology}, 2008, Oxford University Press

\item[] S. Yokoyama and J. Soda, 2008, JCAP, 8, 5
\end{itemize}

\end{document}